\documentclass[twocolumn,aps,prl,groupedaddress,amsmath,amssymb,floatfix, showkeys,superscriptaddress,nofootinbib,nobalancelastpage]{revtex4-2}

\usepackage{xcolor}
\usepackage{dcolumn}  % Align table columns on decimal point
\usepackage{bm}       % bold math
\usepackage{datetime}

\usepackage{graphicx}
\usepackage{bm}
\usepackage{threeparttable}
\usepackage[normalem]{ulem}
\usepackage{array,multirow}
\usepackage{appendix}
\usepackage{epstopdf}
\usepackage{booktabs}
\usepackage{natbib}
\usepackage{feynmp}
\usepackage{threeparttable}
\usepackage{CJKutf8}
\definecolor{TITLECOL}{rgb}{0.1,0.2,0.7} %title color
\definecolor{PCOL}{rgb}{0.5,0.06,0.01} %title color
\definecolor{CHAPCOL}{rgb}{0,0.48,0} %chapter color
\definecolor{SECOL}{rgb}{0.1,0.2,0.7} %sec color
\definecolor{CONTENTSCOL}{rgb}{0.1,0.2,0.7} %can choose the table of contents title to have same color as sec
\definecolor{SSECOL}{rgb}{0.25,0,0.48} %ssection color
\definecolor{SSSECOL}{rgb}{0.2,0.08,0.53} %subsubsection color  0.2,0.08,0.53
\definecolor{SHDCOL}{rgb}{0.4,0,0} % heading of section color
\definecolor{ITMCOL}{rgb}{0.4,0,0} % bulletted text of item color
\definecolor{EXCOL}{rgb}{0,0.47,0.01} %color of exercises
\definecolor{DEFCOL}{rgb}{0,0.42,0.01} %color in definitions file for definition headings

\definecolor{URLCOL}{rgb}{0,0.17,0.43} %external link color
\definecolor{LINKCOL}{rgb}{0.05,0.4,0} %internal link color
\definecolor{CITECOL}{rgb}{0.35,0,0.48} %link to bibliography

\definecolor{ngreen}{rgb}{0,0.48,0}

\renewcommand \thesection{\arabic{section}}

\def\sectable#1{
\addcontentsline{toc}{subsection}{~~Table: \textcolor{SSECOL}{#1}}
\begin{table}[h]
\caption{\bf \textcolor{SSECOL}{#1}}
}

\def\bea{\begin{eqnarray}}
\def\eea{\end{eqnarray}}
\def\ben{\begin{equation}}
\def\een{\end{equation}}
\def\benu{\begin{enumerate}}
\def\enu{\end{enumerate}}

\def\bei{\begin{itemize}}
\def\eei{\end{itemize}}
\def\beit{\begin{itemize}}
\def\eit{\end{itemize}}
\def\benu{\begin{enumerate}}
\def\enu{\end{enumerate}}

\def\n{n}

\def\sss{\scriptscriptstyle\rm}

\def\l{^\lambda}

\def\1var{(\bx_1...\bx\N)}

\def\half{\frac{1}{2}}

\def\br{{\bf r}}

\def\bu{{\bf u}}
\def\bx{{x}}

\def\x{_{\sss X}}
\def\c{_{\sss C}}
\def\s{_{\sss S}}
\def\xc{_{\sss XC}}
\def\Hx{_{\sss HX}}
\def\Hxc{_{\sss HXC}}

\def\N{_{\sss N}}

\def\LDA{^{\rm LDA}}

\def\I{^{\rm I}}

\def\sph_int{ {\int d^3 r}}

\def\intr{\int d^3r\,}
\def\intrp{\int d^3r'\,}

\definecolor{SPECOL}{rgb}{0,0.47,0.01}
\definecolor{QUOCOL}{rgb}{0,0,0.2}
\definecolor{SHDCOLb}{rgb}{0.69,0.4,0.1}%{0.01,0.4,0} % heading of section color

\definecolor{SPEQ}{rgb}{0.01,0.4,0.05} %

\definecolor{SPEQv}{rgb}{0.45,0.05,0.45} %

\definecolor{SPEQb}{rgb}{0.01,0.1,0.65} %

\definecolor{SPEQr}{rgb}{0.57,0.05,0.1} %

\def\special#1{{\color{SPECOL}{\bf #1}}}
\def\break{\vskip 0.5cm}

\def\bay{\begin{array}}
\def\eay{\end{array}}
\def\bit{\begin{itemize}}
\def\beit{\begin{itemize}}
\def\eit{\end{itemize}}

\def\ln{\text{ln} }

\def\intr{ {\int d^3 r\,}}

\def\dd{~ \rotatebox{320}{\hspace{-5pt}\vbox to 5 pt {\hspace{-5pt} \hbox to 5pt {$\cdots$}}}\!\! }
\def\sc{^{\rm sc}}

\definecolor{Green}{rgb}{0.016,0.627,0}
\definecolor{Plum}{rgb}{0.17,0,0.45}
\definecolor{LBlue}{rgb}{0,0.34,0.45}
\definecolor{Sepia}{rgb}{0.37,0.17,0.02}
\definecolor{BurntOrange}{rgb}{0.78,0.39,0}

\usepackage{amsmath}
\usepackage{physics}
\usepackage{amssymb}
\usepackage{hyperref}
\graphicspath{{figures/}}
\usepackage{float}

\def\ncpr{\tilde\n_{\br}}
\def\ncprl{\ncpr^\lambda}
\def\ncprlp{\tilde\n_{\br'}^\lambda}
\def\phicprl{\tilde\phi_{\br}^\lambda}
\def\ncprlx{\ncpr^{\lambda=0}}

\def\cpr{_{\br}}
\def\cprl{\cpr^\lambda}

\def\vscp{\tilde v\s^\lambda(\br'|\br)}
\def\vscpp{\tilde v\s^\lambda(\br|\br')}
\def\tPsi{\tilde \Psi_{\br}}

\let\c\undefined
\def\c{_{\sss C}}
\let\l\undefined
\def\l{^\lambda}
\let\xc\undefined
\def\xc{_{\sss XC}}
\renewcommand{\ln}{\rm{ln}}

\begin{document}
\sf

\title{Bypassing the energy functional in density functional theory: Direct calculation of electronic energies from conditional probability densities}

\author{Ryan J. McCarty}
\affiliation{Department of Chemistry, University of California, Irvine, CA, 92697}

\author{Dennis Perchak}
\affiliation{Department of Chemistry, University of California, Irvine, CA, 92697}

\author{Ryan Pederson}
\affiliation{Department of Physics and Astronomy, University of California, Irvine, CA, 92697}

\author{Robert Evans}
\affiliation{H H Wills Physics Laboratory, University of Bristol, Bristol BS8 1TL, UK}

\author{Yiheng Qiu}
\affiliation{Department of Physics and Astronomy, University of California, Irvine, CA, 92697}

\author{Steven R. White}
\affiliation{Department of Physics and Astronomy, University of California, Irvine, CA, 92697}

\author{Kieron Burke}
\email{kieron@uci.edu}
\affiliation{Department of Chemistry, University of California, Irvine, CA, 92697}
\affiliation{Department of Physics and Astronomy, University of California, Irvine, CA, 92697}

\date{\today}

\begin{abstract}
Density functional calculations can fail for want of an accurate exchange-correlation approximation. The energy can instead be extracted from a sequence of density functional calculations of conditional probabilities (CP-DFT). Simple CP approximations yield usefully accurate results for  two-electron ions, the hydrogen dimer, and the uniform gas at all temperatures. CP-DFT has no self-interaction error for one electron, and correctly dissociates H2, both major challenges. For warm dense matter, classical CP-DFT calculations can overcome the convergence problems of Kohn-Sham DFT.
\end{abstract}

\maketitle

Modern electronic structure calculations usually focus on finding accurate ground-state
energies, as many predicted properties of a molecule or a material depend on this ability~\cite{B12}.
Wavefunction-based methods, such as coupled-cluster theory~\cite{BM07,C66}  or quantum Monte Carlo (QMC)~\cite{A76,AZL12}, directly yield energies.
Kohn-Sham (KS) density functional theory (DFT)~\cite{KS65} incorporates the many-electron problem into the exchange-correlation (XC) energy, which must be approximated as a functional of spin densities.
Hundreds of XC functionals with distinct approximations are available in standard codes~\cite{LSOM18}, reflecting the tremendous difficulty in finding general, accurate approximations. Recently, KS-DFT at finite temperatures~\cite{M65} has been tremendously successful in simulations of warm dense matter~\cite{GDRT14,BDMZ20}.  However, it inherits all the limitations of ground-state approximations and becomes impossible to converge for very high temperatures~\cite{WC20}.

We propose an alternative to KS-DFT, in which we directly calculate conditional probability densities, from
which the energy can be calculated.  This bypasses all the difficulties of approximating the XC energy.
The electronic pair density can always be written as
\begin{equation}
P(\br,\br')=\n(\br)\,\ncpr(\br'),
\label{ncpr}
\end{equation}
where $\n(\br)$ is the single particle density, and 
$\ncpr(\br')$ is the conditional probability (CP) density of finding an electron at
$\br'$, given an electron at $\br$.  The standard exact KS potential of DFT, $v\s[\n](\br)$,
is defined to yield $\n(\br)$ in an effective fermionic non-interacting problem~\cite{DG90}.
A conditional probability KS potential (CPKS), $v\s[\ncpr](\br')$ yields $\ncpr(\br')$ from such a KS calculation
with $N-1$ electrons.  Because standard KS-DFT calculations usually yield accurate densities~\cite{KSB11}, an accurate CPKS potential should yield accurate XC energies. Unlike XC approximations built on theories of the XC hole~\cite{PBE96}, here we {\em calculate} that hole.

\begin{figure}[htb]
\includegraphics[width=8.6 cm]{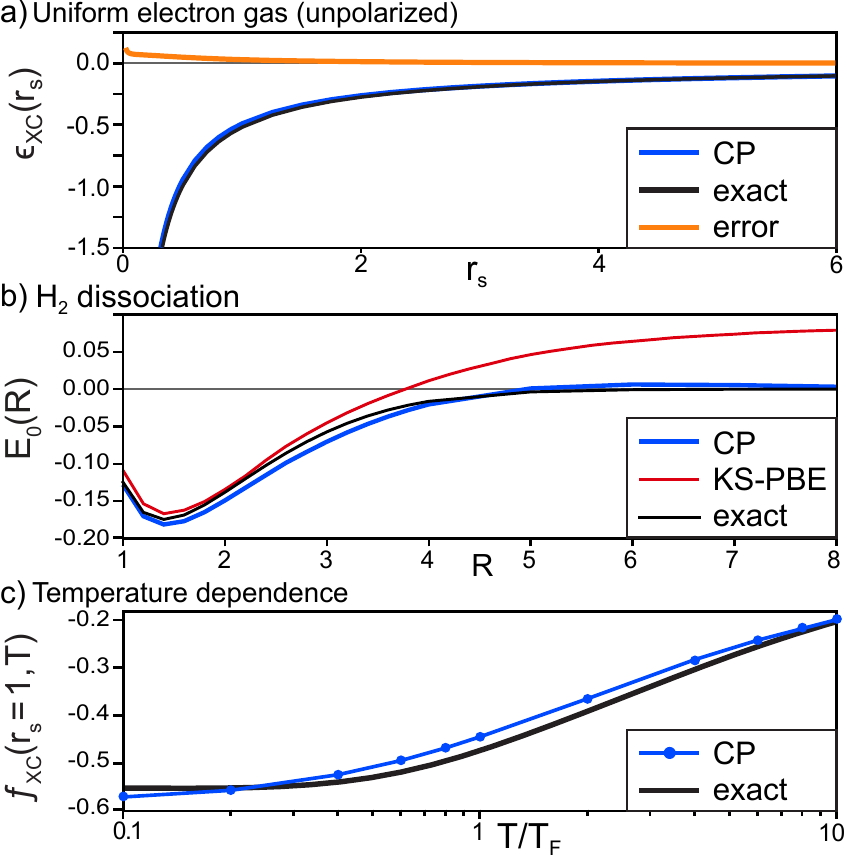}
\caption{CP (blue) and exact (black): (a) XC energy per particle in uniform gas at increasing Wigner–Seitz radii ($r_s$) and $T = 0$, (b) binding energy curve for H$_2$ (red is KS-DFT using PBE~\cite{PBE96}), and
(c) XC free energy per particle at $r\s=1$ as a function of reduced temperature
($T_F$ is the Fermi temperature). Exact from Ref.~\cite{PW92} in (a), Ref.~\cite{GDSMFB17} in (c). Hartree
atomic units used throughout.}
\label{blue_fig}
\end{figure}

Just as in traditional DFT, we construct a simple, universal approximation for the CPKS potential from exact conditions and the uniform gas.
At large separations or high temperatures, the CP potential reduces to adding $1/|\br-\br'|$ to the external potential, as if the missing electron were classical.
We call this a blue electron (i.e. distinguishable from all others), recalling the Percus test particle of classical statistical mechanics~\cite{P62}.
At small separations, the electron-electron cusp condition~\cite{BPL94} requires adding only 1/2 this potential (due to the reduced mass).  We locally interpolate between these two universal limits with representative results shown in Fig~\ref{blue_fig}.
For the uniform gas at zero temperature, our CP potential interpolation is extremely accurate. We added a strong repulsion for $r\s < 1$, to recover the exchange limit.
Panel (b) shows the H$_2$ binding curve, where the inclusion of the electron-electron cusp is vital.  
Unlike semi-local DFT, CP-DFT dissociates the molecule correctly, remaining
spin-unpolarized throughout.
Panel (c) shows CPKS calculations for many temperatures, where the error never exceeds 20\%.
We show later that orbital-free Thomas-Fermi, and even classical, CP calculations agree reasonably with CPKS, are accurate for all $T>T_{F}$, and have errors that vanish in the high temperature limit,
providing an inexpensive alternative when temperatures are beyond the convergence limit of KS-DFT.

\noindent
{\em Theory:} 
We consider non-relativistic purely electronic problems, and use Hartree
atomic units throughout.  The pair density of the
exact ground-state wavefunction $\Psi\l$:
\begin{equation}
P\l(\br_1,\br_2)= N(N-1)\sum_{\sigma_1\sigma_2} \int\,d3\dots\,dN|\Psi\l(1\dots\,N)|^2,
\label{Pldef}
\end{equation}
where $N$ is the number of electrons.
Here $1$ denotes both $\br_1$ and $\sigma_1$, the spatial and spin indices.  
The $\lambda$-dependence is the coupling constant in KS DFT, where the repulsion is
multiplied by $\lambda$ but the one-body potential $v\l(\br)$ is adjusted to keep the
ground-state density $\n(\br)$ fixed~\cite{LP75}.  The XC energy is:
\begin{equation}
E\xc=\half\int_0^1 d\lambda\intr\intrp\frac{\n(\br)\,[\ncprl(\br')-\n(\br)]}{|\br-\br'|},
\label{Exc}
\end{equation}
with $\ncprl(\br')-\n(\br)$ being the $\lambda$-dependent XC hole, defined via the $\lambda$-dependent generalization of Eq. ~\ref{ncpr}.  Setting $\lambda=1$ in Eq.~\ref{Exc} yields $U\xc$, the potential contribution to XC. The integral over $\lambda$ is called the adiabatic connection.

Denote $v\l[\n](\br)$ as the one body potential that yields the unique ground-state density for electron repulsion $\lambda/|\br-\br'|$.  The conditional probability potential is
\begin{equation}
\tilde{v}^{\lambda}(\br'|\br)=v[\ncprl](\br')=v^\lambda[\n](\br')+\Delta\,\tilde{v}^{\lambda}_\br[\n](\br'),
\label{CPv}
\end{equation}
being the unique potential whose ground-state density for Coulomb interacting electrons yields the exact $\lambda$-dependent CP density.
The CPKS potential is found self-consistently:
\begin{equation}
\vscp=v\s[\ncprl](\br')=\tilde{v}^{\lambda}(\br'|\br)+v\Hxc[\ncprl](\br'),
\label{CPKSv}
\end{equation}
where $v\Hxc$ is the Hartree-XC potential~\cite{B12}.  Knowledge of the CP correction
potential, $\Delta\tilde{v}^{\lambda}_\br[\n](\br')$ in Eq.~\ref{CPv},
allows a self-consistent 
KS calculation for the exact CP density.
Uniqueness of the CP potential is guaranteed by the HK theorem. As $\tilde{\n}\cprl(\br')$ is non-negative, normalized to $N-1$, and found from a wavefunction, it is in the standard space of densities, for which we routinely assume KS potentials exist~\cite{L79,L83}.

The above equations are for pure density functionals, and their analogs for spin-density functionals are straightforward (but cumbersome).
Decades of research in DFT can be applied to the study of CP densities and potentials, yielding many exact conditions.  
For example, at $\lambda=0$ where the exchange hole is never positive,
\begin{equation}
    \ncprlx(\br')\leq\,n(\br').
\end{equation}
The CP densities satisfy a complementarity principle:
\begin{equation}
    \ncprl(\br')=\frac{n(\br')}{\n(\br)}\,\ncprlp(\br),
\end{equation}
which is Bayesian, and may be amenable to modern machine-learning methods.
The electron coalescence cusp condition requires
\begin{equation}
    \frac{\partial\ncprl(\br,u)}{\partial u} \bigg|_{u=0}=\lambda\,\ncprl(\br),
\label{eq: exact n CP cusp condition}
\end{equation}
where $\bu = \br' - \br$ and the left-hand side has been spherically averaged over $\br+\bu$~\cite{BPE98}.
Using Ref.~\cite{LPS84}, write
\begin{equation}
\Psi\l(1\dots N)={\sqrt{\frac{\n(\br_1)}{N}}}\,\tPsi\l(2\dots\,N),
\label{tildePsi}
\end{equation}
where $\tPsi\l$ is not antisymmetric
under interchange of the electrons, but is uniquely defined by  Eq.~\ref{tildePsi},
and $\tilde \n\cprl$ is its density.
For large $r$,  Ref.~\cite{LPS84} shows that $\tPsi\l$ becomes a ground-state of the $N-1$ particle system and its gradients with respect to $\br$ vanish, yielding
\begin{equation}
\Delta\tilde\,v^\lambda_{\br\rightarrow\infty}(\br')\to\frac{\lambda}{|\br-\br'|},
\end{equation}
i.e., the blue electron approximation becomes exact in this limit.

For $N=1$, $\tilde \n\cprl(\br')=0$, there is no
self-interaction error~\cite{PZ81}.
If $N=2$, the CP density has just one electron:
\begin{equation}
    \phicprl(\br')=\sqrt{\ncprl(\br')}=\sqrt{\frac{2}{n(\br)}}\,\Psi^{\lambda}(\br,\br')\,,   
\end{equation}
yielding
\begin{equation}
   \vscp-\epsilon^\lambda_{\br}=\half\frac{\nabla'^2\Psi^{\lambda}(\br,\br')}{\Psi^{\lambda}(\br,\br')},
\end{equation}
where $\epsilon^\lambda_{\br}$ is the eigenvalue of the CPKS potential.
Because the wavefunction satisfies the Schr\"odinger equation, we find
\begin{equation}
    \Delta\vscp+\Delta\vscpp=\frac{\lambda}{|\br-\br'|}-E^{\lambda},
\end{equation}
where $\Delta \vscp= \vscp - v^\lambda[\n](\br')
 - \epsilon^\lambda_{\br}$.

\noindent
{\em Approximations:} To perform a CP-DFT calculation, we need a general-purpose approximation to the CP potential, $\Delta\tilde v^\lambda_{\br}(\br')$.  At large separations, the CP potential is simply $\lambda/|\br-\br'|$
for all systems.   At small separations, it is $\lambda/(2|\br-\br'|)$, to satisfy the electron-electron cusp condition,
for all systems.  
We interpolate between these two with a simple local
density approximation
\begin{equation}
    \Delta\tilde\,v^\lambda_\br[\n](\br')\approx\frac{\lambda}{2|\br-\br'|}(1+{\rm\,Erf}\left(\frac{|\br-\br'|}{r_s(\n(\br))}\right)),
\label{eq:vcprl_Erf}
\end{equation}
where $r_s=(3/(4\pi\n))^{1/3}$ is the Wigner-Seitz radius at the reference point.
We use this approximation for all ground-state CP calculations in the paper.
Fig.~\ref{blue_fig}(a) and \ref{blue_fig}(b) use Eq.~\ref{eq:vcprl_Erf} combined with standard DFT
approximations for $v\xc$. Fig.~\ref{blue_fig}(c) uses simply $\lambda/|\br-\br'|$, 
as the difference is negligible at significant temperatures.

\noindent
{\em Uniform electron gas:}
The $N$-electron density is trivially
a constant, and the one-body potential vanishes.
The CP calculation is for $N-1$ electrons in a KS potential:
\begin{equation}
v\s(r)=\Delta\tilde\,v(r)+\int\,d^3r'\frac{\tilde \n(\br')-\n_0}{|\br-\br'|}+v\xc\LDA[\tilde\n](\br),
\label{unifvs}
\end{equation}
where $\n_0=N/V$ and
\begin{equation}
   \Delta \tilde v(r) = \Delta\tilde v^{(\lambda=1)}_{\bf 0}(n_0,r) + A(r_s) e^{-r^2/2 \sigma(r_s)^2}.
\end{equation}
The second term is added to recover the
correct high-density limit, i.e., the simple $n^{4/3}$
exchange energy. By calculating many $r_s$ values we can integrate over $r_s$ to perform the adiabatic connection with only $\lambda=1$.  The XC potential is from~\cite{VWN80}. The strength and range parameters of the added Gaussian potential are fitted to~\cite{VWN80} for $r_s = 0.02$, where exchange dominates.
The density is found self-consistently in a sphere using Fermi-surface smearing ($T=0.05 T_{F}$) and $N=512$. Imposing zero density flux through the surface of the sphere minimizes boundary effects. Further details will appear in a forthcoming paper.

\begin{figure}[htb]
\includegraphics[width = 8.6 cm]{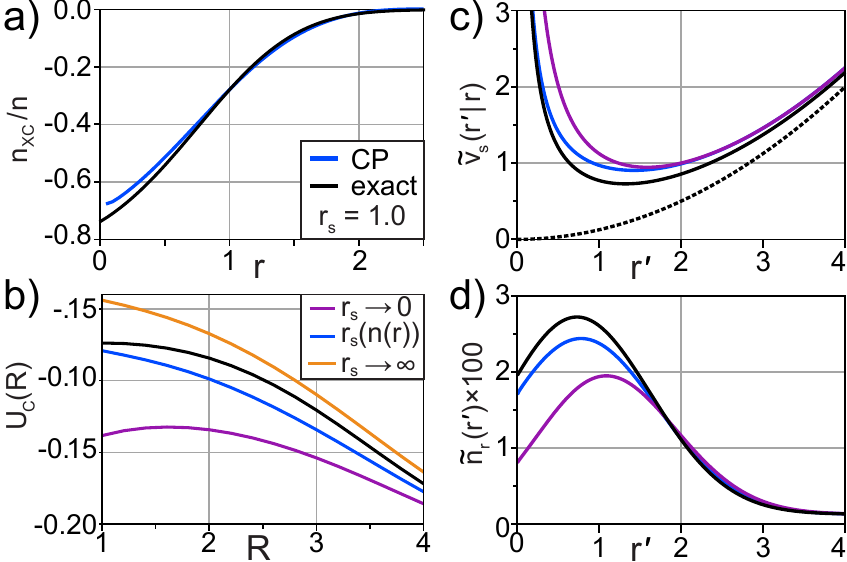}
\caption{
Pure blue electron approximation (purple), half pure blue approximation (orange), interpolation (Eq.~\ref{eq:vcprl_Erf}, blue), and exact (black):
(a) normalized XC hole densities for the uniform gas at $r_s=1$ with exact parameterization from Ref.~\cite{PW92b},
(b) $U_c(R)$ from H$_2$, using exact $n(r)$ and producing errors below $20\%$,
(c) Hooke's atom $\tilde v\s(\br'|\br)$, with the external potential $r^{\prime\,2}/8$ (black dashed), and (d) Hooke's atom $\ncpr(r')$.
}
\label{fig:merged2_4}  
\end{figure}

Fig.~\ref{fig:merged2_4}(a) compares the hole density to the parameterization of the uniform gas XC hole~\cite{PW92b}.  The agreement is very good, with the lowest accuracy from the on-top region, which minimally affects the XC energy.

\noindent
{\em Atoms and molecules:}  
We applied Eq.~\ref{eq:vcprl_Erf} to highly accurate calculations of 2-electron systems. These calculations were done using a new type of basis function called gausslets~\cite{W17,WS19} which are tailored for density matrix renormalization group calculations~\cite{W93} and based on wavelets.
Gausslets resemble a variable-spaced real-space grid. The two-electron Hamiltonian terms have only two indices, $V_{ij}$, unlike the four indices needed in a standard basis.  The grid-like structure make CP calculations easy to implement.  A blue electron sitting at a point in space sits on a gausslet, $i$, located at its reference, ${\bf r}_i$. The repulsive one-electron potential at $i$ is simply row $i$ of $V_{ij}$, and integration likewise becomes point-wise sums. Recent innovations add a Gaussian basis to better describe atomic core behavior, further described in a forthcoming work. We used $2000$ or less gausslets with total energies errors below $0.1$ mH for $Z=1$ and $Z=2$. To find the conditional probability using  Eq.~\ref{eq:vcprl_Erf}, we find the ground state of an $N\times N$ matrix with the Lanczos algorithm~\cite{Lanczos50} and repeat $N$ times. Gausslets make an excellent basis for CP calculations, but in any basis, CP calculations are receptive to parallel computing, as each value of $\bf{r}$ and $\lambda$ can be computed independently.

Accurate densities from standard DFT calculations are needed for CP calculations.
For 2-electron ions presented in Table~\ref{tab: two electron ions}, we choose Hartree-Fock, as it provides a bound density even for H$^-$~\cite{cox2020bound}.  We performed the double integral over $\br$ and $\br'$ to find the potential contribution to correlation, $U\c$. The virial theorem for atoms (relating the total energy to total kinetic energy, $E=-T$) then allows us to deduce $E\c$. 
For He, the ground-state energy error is -6 mH, while that of PBE is +10mH.  As $Z\to\infty$, the CP calculation correctly yields a finite value.  At $Z=1$, the error has increased to 10mH, but H$^-$ is not even bound in a KS-DFT calculation with standard approximations ~\cite{KSB11}.

\begin{table}[ht]
\begin{center}
\renewcommand{\arraystretch}{1.4}
\begin{tabular}{|c|c|c|c|c|c|c|} 
\hline
$Z$ & $E\x^{\text{HF}}$ & $V^{\text{CP}}_{\text{ee}}$ & $U^{\text{CP}}\c$ & $U^{\text{Exact}}\c$ & virial $E^{\text{CP}}\c$ & $E^{\text{Exact}}\c$ \\
\hline
1.0 & -0.3959 & 0.2918 & -0.1041 & -0.0698 & -0.0523 & -0.0420 \\
\hline
2.0 & -1.0257 & 0.9301 & -0.0956 & -0.0786 & -0.0479 & -0.0421 \\
\hline
3.0 & -1.6516 & 1.5521 & -0.0995 & -0.0832 & -0.0504 & -0.0435 \\
\hline
4.0 & -2.2770 & 2.1750 & -0.1020 & -0.0857 & -0.0525 & -0.0443 \\
\hline
6.0 & -3.5273 & 3.4226 & -0.1047 & -0.0881 & -0.0563 & -0.0452 \\
\hline
\end{tabular}
\caption{Results for 2-electron Helium-like ions using HF densities. Virial $E^{CP}_c$ is derived from the virial theorem for atoms.}
\label{tab: two electron ions}
\end{center}
\end{table}

The virial trick only works for Coulomb-interacting atoms and molecules at equilibrium. Otherwise, we need to perform the adiabatic connection integral. For $N=2$, we know the exact result as $\lambda\to 0$ (exchange limit),
where $\ncprlx(\br')=\n(\br')/2$. By definition, for 2-electrons we have  
\begin{equation}
\vscp = v\s[\n](\br') -\lambda v\Hx[\n](\br') - v\c^\lambda[\n](\br') + \Delta \vscp \, .
\label{eq:adiabatic connection exact}
\end{equation}
In practice, obtaining $v\c^\lambda[\n](\br')$ is difficult, and we approximate
\begin{equation}
    \vscp \approx 
    \begin{cases} 
    v\s[\n](\br') \, , \, \lambda = 0 \\
    v[\n](\br') + (1-\lambda) v\Hx[\n](\br') + \Delta \vscp  \\
    \end{cases}
\label{eq:adiabatic connection weak correlation}
\end{equation}
to recover the exchange limit exactly. In the following calculation for H$_2$, we utilize the interpolated blue approximation, Eq.~\ref{eq:vcprl_Erf}, for $\Delta \vscp$ and the exact density $n(\br')$ throughout. We run for $\lambda \in \{0.0,0.1,0.3,0.5,0.7,1.0 \}$, and fit to a first-order Pad\'e approximant, which is integrated analytically.

\begin{table}[ht]
\begin{center}
\renewcommand{\arraystretch}{1.4}
\begin{tabular}{|c|c|c|c|c|c|c|} 
\hline
$R$ & $E\x$ & $V^{\text{Blue}}_{\text{ee}}$ & $U^{\text{Blue}}\c$ & $U^{\text{Exact}}\c$ & $E^{\text{Blue}}\c$ & $E^{\text{Exact}}\c$ \\
\hline
1.0 & -0.7472 & 0.6688 & -0.0785 & -0.0732 & -0.0433 & -0.0400 \\
\hline
2.0 & -0.5698 & 0.4720 & -0.0978 & -0.0835 & -0.0587 & -0.0478 \\
\hline
4.0 & -0.4323 & 0.2576 & -0.1747 & -0.1692 & -0.1359 & -0.1318 \\
\hline
8.0 & -0.3749 & 0.1241 & -0.2497 & -0.2499 & -0.2445 & -0.2477 \\
\hline

\end{tabular}
\caption{H$_2$ energies versus $R$, where $E^{\text{Blue}}\c$ is computed from Eq.~\ref{eq:adiabatic connection weak correlation} with the exact density.}
\label{H2_adiabatic_connection}
\end{center}
\end{table}

The binding curve for H$_2$ as a function of bond length is shown in Fig 1(b), with components given in Table~\ref{H2_adiabatic_connection}.
Fig.~\ref{fig:merged2_4}(b), shows $U\c(R)$ for 3 distinct choices of CP potential.
As $R\to\infty$, any version of the blue electron approximation becomes accurate. Consider what happens as the bond is stretched.  The exact
wavefunction has Heitler-London~\cite{HL27} form:
\begin{equation}
\Psi\l(\br_1,\br_2)=\frac{1}{\sqrt{2}}\left( \phi_A(\br_1)\, \phi_B(\br_2) + \phi_B (\br_1)\, \phi_A(\br_2) \right)
\label{HL}
\end{equation}
where $\phi_A$ and $\phi_B$ are atomic orbitals localized on each of the two protons.
This yields a conditional density:
\begin{equation}
\n\cprl(\br')=\n_B(\br'),~~~~~\br~{\rm near}~A
\end{equation}
and vice versa,
for all $\lambda \neq 0$.  Thus the Coulomb energy of the pair density vanishes
due to the lack of overlap, and each atomic region correctly yields a one-electron energy
of a separate hydrogen atom.
Standard semilocal DFT
must choose between retaining the correct spin symmetry, as in the PBE curve of Fig 1(b), or sacrificing accurate spin densities\cite{PSB95}.  At the formal level,
CP-DFT is an exact theory for bond dissociation, unlike the on-top hole theory of Ref.~\cite{PSB95}. 

Hooke's atom consists of two Coulomb repelling electrons in a harmonic potential of force constant $k$~\cite{kais1993density}. At $k=1/4$, the density is known analytically, and at $r = 0$, the exact $\vscp$ is radial. In Fig.~\ref{fig:merged2_4}(c) and ~\ref{fig:merged2_4}(d) we compare the blue electron approximation, our interpolation formula Eq.~\ref{eq:vcprl_Erf}, and the exact CP potential and the resulting densities $\ncprl(r')$.
Note the accuracy of the blue approximation for large $r'$, and the cusps as $r'\rightarrow\,r$ in the exact and approximate CP densities.

In practical calculations, one does not have access to exact densities, but usually KS-DFT
densities from standard approximations are accurate, and in many cases where they are not,
Hartree-Fock densities are better\cite{KSB13}.  In principle, if neither suffices, densities
could be found self-consistently by minimizing the energy from CP calculations with respect
to the $N$-electron density.

\noindent
{\em Finite temperatures:}  
Possibly, the most important application of CP-DFT is for thermal equilibrium in warm dense matter~\cite{GDRT14}. While thermal KS-DFT calculations have been very successful,
finding consistent temperature-dependent approximations is more difficult than at
zero temperature~\cite{DGB18}.  Moreover, calculations using KS solvers eventually fail at extremely high temperatures, due to convergence difficulties with orbital sums.

For finite temperatures, Eq~\ref{Exc} translates to $F_{XC}$, the XC contribution to the
Helmholtz free energy, which includes entropic contributions~\cite{M65,PPFS11}.
To find accurate
CP densities, we solve the KS equations with finite temperature occupations.
(Thermal corrections to $v\xc$ are argued to have little effect on the orbitals~\cite{SPB16}). Fig. 1(c) shows results for the potential XC free energy at $r_s = 1.0$ for a wide range of temperatures. The black curve displays the analytical parameterization (Ref.~\cite{GDSMFB17}). The CPKS approximation mildly overestimates $f\xc$ for $t= T/T_{F}$ between about 0.2 and 9.
This accuracy has been achieved from our trivial CPKS
calculation, without any quantum Monte Carlo or other many-body solver.

But for high temperatures, KS-DFT calculations fail to converge due to the exponential growth in orbitals that contribute, and our calculation is no exception.
We therefore performed a much simpler CP calculation using the Thomas-Fermi (TF) approximation~\cite{T27,F28}, often employed in plasma physics ~\cite{FMT49, L17}, and implementing the simple blue approximation. We first solved the TF equation at $T=0$ to initiate iterations for a full numerical solution. We make a simple interpolation of Perrot's~\cite{P79b} accurate parameterization of the Helmholtz free energy density $f_0 (n)$ of the uniform non-interacting electron gas constructed to yield the correct $T=0$ and (classical) $T\rightarrow\infty$ limits:
\begin{equation}
f_0 (n) = k_B Tn \left(\ln(y)-c+ay^{\frac{2}{3}}\right),
\label{CLapx}
\end{equation}
 where $y = \pi^2 n/ \sqrt{2} (k_{B}T)^{3/2}$, $c=1-\ln(2/{\sqrt{\pi}})$, and $a=9(2/3)^{1/3}/10$. The Fermi temperature is given by $k_{B}T_{F}=(3\pi^{2}n)^{2/3}/2$. As $T \to 0$, $f_{0}(n) = 3nk_{B}T_{F}/5$ as required. TF theory corresponds to minimizing the Mermin~\cite{M65} grand potential functional ignoring XC and making the local density approximation $F[n]= \int d^{3}rf_{0} (n(r))$ for the non-interacting Helmholtz free energy.

\noindent
{\em Classical connection}: In the classical limit TF theory reduces to the Poisson-Boltzmann (PB) theory used to treat electrical double layers and many other properties of electrolyte solutions and ionic liquids~\cite{HM13}. In the high temperature limit we can ignore the third term in Eq~\ref{CLapx} yielding
\begin{equation}
    F[n] = k_{B}T\int d^{3}r\, n(\mathbf{r})\left(\ln\left({\frac{n(\mathbf{r})\lambda^{3}}{2}}\right)-1\right),
    \label{freeEfun}
 \end{equation}
where $\lambda=(2\pi/k_{B}T)^{1/2}$ is the thermal de Broglie wavelength. Eq.~\ref{freeEfun} is identical to the Helmholtz free energy functional of the ideal classical gas, apart from the residual spin degeneracy factor $(2s+1)$. Employing Eq.~\ref{freeEfun} from the outset corresponds to implementing the classical DFT~\cite{HM13, E79} that generates PB theory for the one-component plasma.  In the classical limit the TF screening length, $\lambda_{TF}$~\cite{AM76}, reduces to the Debye length $\lambda_{D}$ of the OCP, given by
 $(\lambda_{D} )^{-2} = 4\pi e^{2} n/k_{B}T$.

\begin{figure}[htb]
\includegraphics[width = 8.6 cm]{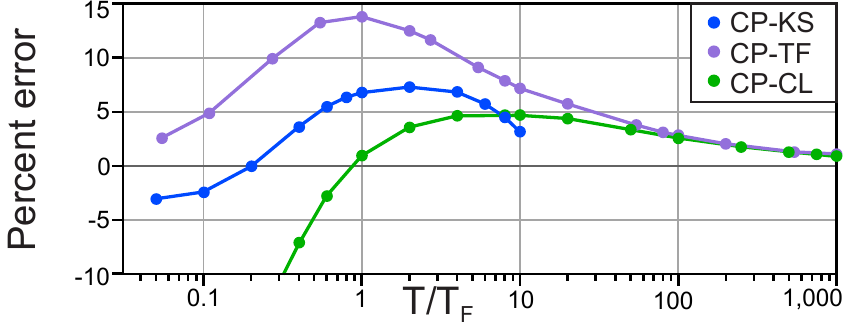}
\caption{Percentage error of uniform gas potential XC free energy per electron for the CP-DFT calculations within KS (blue), TF (purple), and classical (green) approximations relative to the parameterization of Groth et al.~\cite{GDSMFB17}.}
\label{unifT}
\end{figure}

Fig.~\ref{unifT} shows relative XC free energy errors as a function of $t= T/T_{F}$ over a larger temperature range than Fig.~\ref{blue_fig}(c). The blue KS approximation (blue curve) performs well across its range. CP-TF (purple) overestimates up to $t \approx 10$; for larger values, all results merge. The classical approximation (green) becomes exact at sufficiently high $t$.

In the classical limit (Boltzmann statistics) the CP approach is equivalent to the Percus test particle procedure~\cite{C91, P62}.
Fixing a (classical) particle at the origin constitutes an external potential for the others.
The resulting one-body density is proportional to the pair correlation function of the liquid~\cite{ACE17}. The Percus procedure for quantum systems was pioneered by Chihara~\cite{C91} and the most successful applications relate to liquid metals and electron-ion correlations~\cite{AL00}.

Lastly, we mention a connection with factorization schemes in the ground state.
Eq.~\ref{tildePsi} can be used to find a differential equation for $\tPsi\l(2 \dots N)$.
But this is {\em not} an eigenvalue equation
that you solve with given boundary conditions.  Such conditional wavefunctions
are not always the lowest eigenstate if one treats this as an eigenvalue problem~\cite{MAKG14}.
Moreover, the potential experienced by $\tPsi\l(2 \dots N)$ depends on all $N-1$ coordinates,
so it is not amenable to the standard KS treatment.  Thus this seems an unlikely route for
deriving other exact properties.

In conclusion, CP-DFT calculations provide a useful alternative to standard KS-DFT.  While more expensive, they are
highly parallelizable and in important cases, can succeed where KS-DFT often fails.  Most importantly, such
calculations bypass the need to approximate the XC functional and its potential in difficult cases, such as bond
breaking.  Our CP potential approximation becomes exact in many limits. It may be exact even for strictly correlated electrons, where
\begin{equation}
    \ncprl (\br')\to\sum^{N-1}_{j=1}\delta^{(3)}(\br'-{\bf f}_j (\bf r)),
\end{equation}
and ${\bf f}_j (\br)$ is a co-motion function~\cite{GSV09, GVG18}.  Several longer works will follow.

\begin{acknowledgments}
R.J.M. supported by University of California
President's Postdoctoral Fellowship, 
K.B. and D.P. by DOE DE-FG02-08ER46496,
R.P. and S.R.W by DOE DE-SC0008696, and
R.E. by Leverhulme Trust EM 2020-029/4.
K.B. thanks John Perdew for suggesting a variation on this in 1993.
\end{acknowledgments}

\bibliographystyle{apsrev4-2}
\bibliography{Master.bib}

\label{page:end}
\end{document}